\documentclass[journal]{IEEEtran}

\ifCLASSINFOpdf
   \usepackage[pdftex]{graphicx}
   \usepackage{float}
   \graphicspath{{../pdf/}{../jpeg/}}
   \DeclareGraphicsExtensions{.pdf,.jpeg,.png}
\else
\fi

\begin{document}

\title{UAVs as Mobile Base Stations}

\author{Matteo~Strada, University of Trento}

\markboth{Advanced Network Modeling and Design, Research Proposal, AY 19/20}
{Shell \MakeLowercase{\textit{et al.}}: Bare Demo of IEEEtran.cls for IEEE Journals}

\maketitle

\begin{abstract}
The research presented here regards the use of drones as mobile base stations. 
With the advent of fifth generation (5G) RANs, networks will have to handle an increasing amount of diverse devices. This will lead to a greater demand on current telecom companies and infrastructures. Solutions to meet these current new demands might not suffice in guaranteeing the required service for the increasing demand. 
In emergency situations in particular, using terrestrial infrastructures is impossible for various reasons. All these factors suggest exploiting unmanned aerial vehicles as mobile base stations, overcoming all these adversities.
The current state of the art is evaluated, in order to pinpoint the best approach which fulfills all requirements; in doing so, the business impact is also taken into account.
This proposal is a guideline for further research and the eventual creation of a new startup.
The outline of the research proposal is as follows:
\begin{enumerate}
    \item This section is a brief overview of the current state of the art, examining what is readily available and what needs to be further developed.
    \item This section covers the research proposal, an appropriate solution is suggested, overcoming the shortcomings of the current state of the art.
    \item This section covers the possible business impacts and outcomes that this research could have if further developed and implemented.
\end{enumerate}
\end{abstract}

\section{STATE OF THE ART}
\IEEEPARstart{W}{ith} the advent of fifth generation (5G) RANs, networks are required to handle an ever growing amount of diverse devices. All these connections, especially with the rise in IoT device usage, will have different requirements on reliability, latency, etc. compared to current 4G cellular networks. This will determine a heavy demand on current telecom companies and infrastructures. Solutions to meet these new demands have already been researched, such as heterogeneous networks, however these might not be sufficient to cover the increasing demands in all situations. 
In emergency situations, such as disaster response/recovery, the deployment of terrestrial infrastructures is economically unfeasible or impractical. This is why, in this proposal, the idea of leveraging unmanned aerial vehicles (UAVs/drones) as mobile base stations is considered a promising and feasible way to overcome these setbacks.

\subsection{UAVs TECHNOLOGY}
The current state of the art for UAVs feature diverse type of drones depending on their own operational features\cite[p.~1-4]{UAVNetworks}. The two main categories are HAP (high altitude platform) and LAP (low altitude platform). In each of these categories, we can further divide UAVs depending on their physical features, which are, multirotors and fixed wing type aircrafts. As mentioned before, each of these platforms have their advantages, for instance HAPs have longer endurance than LAPs, while LAPs have lower operational cost than HAPs. Furthermore, fixed wings have greater speed and payload than multirotors, while multirotors operate with ease in confined or small environments. We will discuss in particular how to exploit each UAVs type advantage in the proposal section.
In both cases the main concern is energy consumption. Ideally, in an aerial network, in fact, there is the need to stay airborne as long as possible. This is not always achievable and definitely a difficult task, but the literature provides different solutions to do so. First of all, from my personal experience in the field, it is mandatory to correctly power and size the UAVs in order to have the correct power-to-weight ratio, hence not wasting resources that could be used otherwise. This seems trivial, but examples showed that it is not always as obvious as it should be. Second, the correct choice for battery chemical technology should be considered. In fact, nowadays there is a widespread application of LiPo batteries, which have an exceptional stress resistance and power-to-weight ratio, but might not be the best choice for our UAVs. LiFe or LiIon batteries, in fact, definitely provide less stress resistance and a lot less instant current than their LiPo counterparts, but are a much more energy dense, so if appropriately provisioned for the specific equipment, they will outlast any LiPo counterpart. 
On the other hand, a completely different approach could be adopted, as we will see in the proposal section, such as using a multirotor/fixed wing hybrid or even a glow powered multirotor.
All the aforementioned options, though, are examples of small and lightweight UAVs, but we don't want to rule out the possibility of integrating military-grade high-powerer options. In fact, the latest drone from Leonardo, the Falco Xplorer, might check all the boxes for our required characteristics due to its 24 hour of flight time and 350kg of payload\cite{Leonardo}.


\subsection{SDN EXAMPLE TECHNOLOGY}
The other drawback of airborne drones is that they have a limited payload so the base transceiver station (BTS) must be as small as possible. Because of this, current techniques regarding software defined networks (SDN) should be used and exploited as much as possible. An actual real world example of how this can be done is Athonet’s CUBE or BACKPACK\cite{Athonet}. These two technologies are relatively small: the former lighter than 30Kg and circa 12Kg the latter, and therefore transportable. The main idea behind them, as aforementioned, is that of providing a mobile virtual network operator (MVNO) by providing a centralized Mobile Core (EPC/NC) with proven roaming capability to national MNOs. This specific solution can interface with current existing national MNOs, so it definitely guarantees the crucial aspect of the integration between aerial and terrestrial networks. It is not the only suitable solution, since, despite their fairly small dimensions, it can still be too big for a single (or a fleet) of UAVs, hence a hybrid approach could be adopted to achieve our goal.

\subsection{COMMUNICATION BETWEEN DRONES AND TERRESTRIAL BTS}
Finally, the other necessary feature of this research proposal is how to be able to communicate with terrestrial BTS or orbital satellites needed to forward network traffic according to network topology. Wireless communication should be carefully chosen in order to avoid common problems related to wireless network and be capable of working without interfering with UAV’s radio and telemetry up link and down link. In this case, the current state of the art provides enough options and techniques to achieve this purpose.
Strictly speaking of frequencies, these should be carefully chosen also depending on each countries regulations, which might be strict boundaries. It seems that mmWave frequencies are currently the way to go: since they range between 30 and 300GHz, they allow using a frequency space which is free to use in most countries, thus not requiring special licences as of now. These frequencies, though, are pretty sensitive to path loss and losses typical of wireless links. On the upside though, mmWave frequencies allow using lower and hence more robust frequencies for the actual UAV radio link. In fact, the whole 2.4GHz band is available to be exploited for this specific use. This band is in fact extremely common in the field of RC air crafts\cite{RCprotocols}, supporting a wide variety of proprietary protocols based on DSSS/FHSS modulation techniques. We can even lower the frequency further to achieve longer ranges, like what has been done by TBS’s Crossfire and FrSky R9, which operates at 868/910MHz, or going to 433MHz as with the Dragonlink RC or even combining those two techniques, as done by JETI radio systems.

\section{Research Proposal}
\IEEEPARstart{D}{epending} on the actual use case needed, the accurate choice of UAVs should be carried out. Usually, as previously briefly presented, the standard calls for a fixed-wing for high endurance operations and a multirotor for high-flexibility scenarios. My idea is that combining these two aspects would enable network operators to exploit the best of both worlds. 

\subsection{UAV design characteristics}
In order to do this, we could build a long-endurance gasoline powered UAV, like that proposed by Tao Pang, Kemao Peng, Feng Lin and Ben M.\cite{GasMultirotor}. This is definitely a good idea, but I would not proceed this way, since having a gasoline powered motor further complicates the whole system and is certainly less ideal in urban scenarios because of pollution (noise and air). In my opinion, the ideal is combining their variable propeller pitch approach with that of the wing-multirotor hybrid Songbird\cite{Songbird}. This way we can build a variable pitch multirotor which can, depending on the specific use case or weather conditions, transform like what’s done by USAF Ospreys into a flying wing. 

\begin{figure}[H]
\centering
\includegraphics[width=3in,height=2in,keepaspectratio]{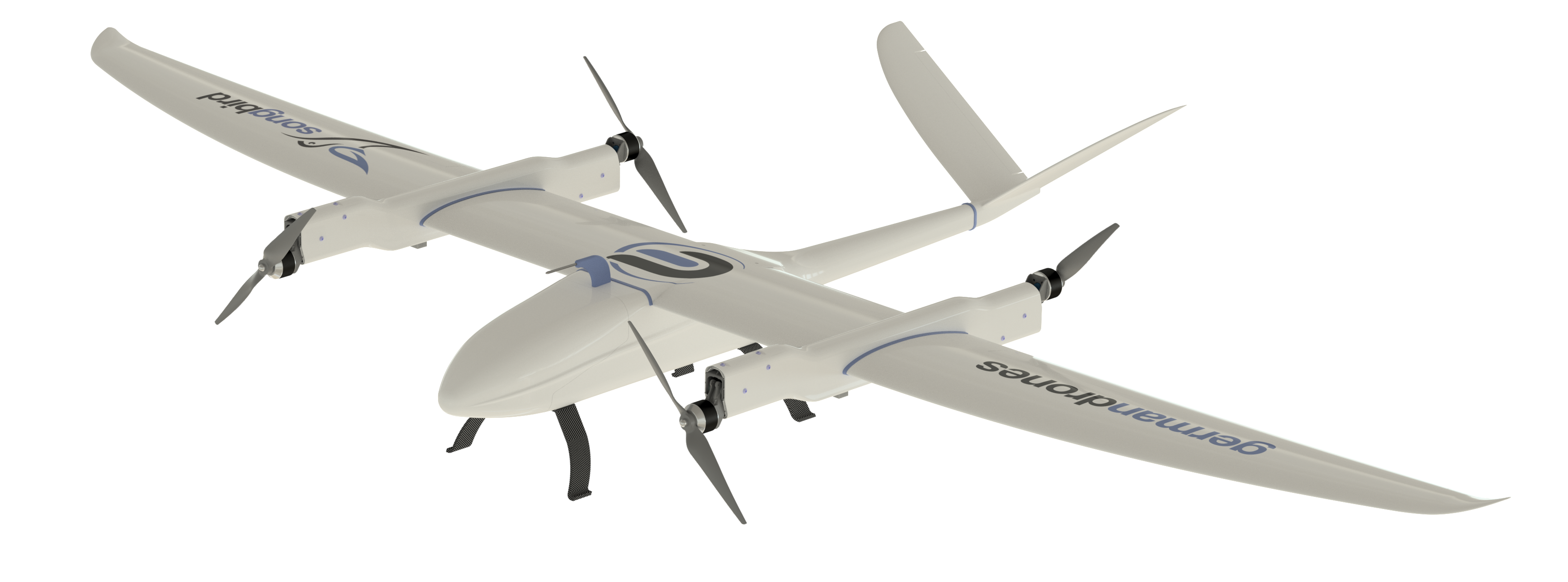}
\label{fig_songbird}
\centering
\caption{The Songbird UAV}
\end{figure}

As design characteristics for this project I would recommend following these guidelines:
\begin{itemize}
    \item the airframe should be made by carbon-fiber, hollow fiberglass structures or other light materials (e.g. titanium/graphene) in order to keep make it as light as possible, maximizing the payload.
    \item the added wing surface can be exploited to fit solar panels to recharge batteries while in flight\cite[p.~24]{UAVComms}.
    \item the motors should have a low kV value, to be able to use highest voltage batteries in order to achieve maximum power with the lowest current consumption, hence having more reliable batteries.
    \item strictly speaking of batteries the UAVs should use LiIon cells, to exploit their higher energy density than standard LiFe and LiPo batteries.
    \item the airframe should have a structure that can glide as much as possible as this can favor huge power savings by exploiting weather conditions such as ascending thermics.
    \item equipping each UAV with an autonomous driving system, such as a small ML-base guidance system.
\end{itemize}

\subsection{UAVs Guidance system}
The use of ML can be exploited for different reasons: to calculate the best and most power efficient flying pattern that can be followed to still guarantee connectivity in the specific area; be able to coordinate a swarm of UAVs in case the payload for the mobile base station is not enough and has to be divided between multiple UAVs and finally to guarantee a “safe” (or the safest) behaviour in case of emergencies or equipment faults.
The UAVs operation will be coordinated by a unique point for their specific area of operation. This point of control corresponds to the terrestrial BTS which will act as a controller both for the drone network and as a relay for the main network. This BTS could also be built like a tower, where UAVs currently not in operation can be stored. This way the UAVs have a single place where they can be stored as backups or where they can automatically land for maintenance or battery recharging. Keeping backups in the (relatively) close vicinity will guarantee continuity of service in case something unexpected happens, and the ability of base stations to handle handoffs will allow for hot swaps of the UAVs when they need maintenance.

\subsection{Network operation}
Now it’s time to consider the technologies needed for the specific network operation. As previously discussed, the use of SDN techniques is fundamental to enable a base station to be carried on UAVs, since it will cut power and weight constraints a lot by virtualizing a network as much as possible. The current state of the art solutions, though, are still too bulky to effectively address the problem, so different approaches should be considered. For instance, instead of fitting a drone with an entire BTS, we could use only (one or more) femtocells. This approach will definitely be the best to develop on drones since its power and payload requirements are easily handled by drones, but still there are all the common limitations due to the use of femtocells and the need of a common BTS. These limitations however can be overcome by extending the transmission ranges of the femtocells with the appropriate techniques (to be discussed later) and by having the correct number of femtocells to allow for the appropriate coverage. 
Another option, since there is not actually the need of creating a full-fledged network only with drones, is to have the UAVs acting as SDN switches on data plane for collecting context information in a distributed way, while the ground BSs (or even the satellites BS) are controllers gathering data and making control decisions on network functions and resource allocation. Helped by SDN, network reconfiguration and resource allocation among a swarm of UAVs can be conducted in a more flexible way. 
In both these hypothesis UAVs will act as a relay between the UEs and the “main” BTS, which (in these case) is still terrestrial, but will be smaller and cheaper for establishing the infrastructure, since it relies on the UAVs to handle part of the computation and network operation.

\subsection{Transmission techniques}
Finally, to be able to carry out the task proposed in the aforementioned approaches, it is necessary to extend and reinforce the transmission ranges and techniques. In particular, the use of mmWave frequencies to communicate between the UAV and the UEs needs some additional measures, because of the nature of mmWave frequencies\cite[p.~16]{UAVComms}. Being high frequencies and short wavelengths, in fact, means that they are extremely susceptible to problematics like propagation loss. A possible solution to this is using a combination of beamforming techniques, which can be exploited to construct a narrow directional beam and overcome the high path loss or additional losses caused by atmospheric absorption and scattering. The fact that on a UAV network, the BTS is moving further complicates this task and calls for a better approach. Beamforming should in fact be made more efficient by combining tracking equipment so that the transmitter and receiver can move with a greater degree of freedom, hence guaranteeing that their movement is always relative to the position of the user's equipment, thus making micro corrections in correlation with UAV movements. Different techniques have been studied, relying on measures of link quality, like RSSI, to properly adapt to movement. Furthermore I think that preparing different antenna combinations (e.g. arrays of microstrip + horn + helix) could also be an option, so that the airborne BTS can also dynamically change the actual hardware transmitting to be able to better beamform and carry out its task.

\section{BUSINESS ANALYSIS}
\IEEEPARstart{T}{he} possible applications of UAVs networks are numerous and varied.\cite[p.~34]{UAVComms}\cite[p.~5-8]{UAVNetworks} 

\subsection{Disaster Response and Recovery}
The first and most useful is that of a rapid network deployment in case of natural disasters or terrorist attacks. Studies have shown in fact how, after some large scale natural events, the network continuity was disrupted. Taking hurricane Katrina as an example, the aftermath showed that approximately 25\% of subnets were inferred as unreachable, and while 62\% of unreachable subnets were small subnets at the edges of networks, the majority (73\%) of unreachable subnets lasted longer than four weeks. This is even more evident, showing how a natural disaster like Katrina causes extreme damage on networks, but is even more indicative on how long it took to effectively recover the normal network functionality. Since network-service disruption is inevitable in these extreme cases, it is crucial to have effective techniques to overcome this. Guaranteeing network continuity is imperative nowadays: system downtime results in loss of revenue that might even bring a business venture to the brink of bankruptcy\cite{SystemDowntime}. Small business especially rely every day more and more on cloud-based applications, so the loss or excessive sluggishness of network connectivity can cripple their entire operations, while bigger businesses, like Amazon/Google/Facebook, will suffer a lot more from the longer-term effect of brand damage, like what happened in the earliest stages of AWS, when for different reasons, the company produced an uptime of 99.99\%, 52.56 minutes of system downtime over the course of a year, which, given today standards, is unacceptable for many customers. Finally the possible data loss and/or exposure to attacks resulting from network continuity problems is something that is a threat to all business ventures. Nowadays, in an increasingly interconnected world, the impact of network-service disruption goes beyond a strictly business perspective, and can in fact result in heavy losses of human lives. Having a reliable network becomes crucial to coordinating rescue operations in case of natural disasters, so, being able to quickly re-instate the lost infrastructures using a UAV network to provide coverage will definitely prove valuable for rescuing people in distress.

\subsection{Network Establishment Feasibility Study}
Another interesting application of UAVs established network is that of studying the feasibility and profitability of a terrestrial network. This application is particularly useful in rural areas where the “standard” network infrastructure is not present or is not able to guarantee coverage. If, in fact, a network operator decides to expand its current network, it could use a UAV network as a temporary solution to probe the specific area and further model if establishing a proper terrestrial network infrastructure in the specific area makes sense both business and cost wise. Plus, using this approach to probe the area of interest could also further be explored and expanded in case in that area there is a business interest but establishing a terrestrial network would be too costly. In this case, in fact, the UAVs network could be kept there for the time being, thus providing coverage, and only once the area is ready and big enough to build a terrestrial network, switch to it.

\begin{figure}[H]
\centering
\includegraphics[width=0.4\textwidth]{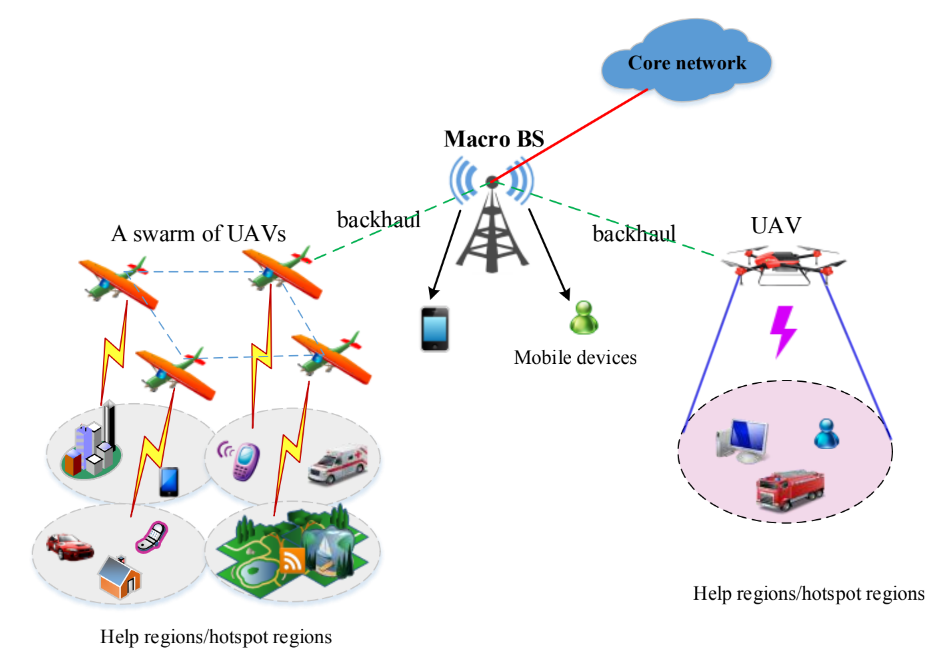}
\label{fig_target_specific_serving}
\centering
\caption{UAVs as aerial BSs serving a target area}
\end{figure}

\subsection{Traffic Offloading}
One of the most useful applications of mobile aerial networks is that of providing traffic offloading for already established networks. Thinking, for instance, of large events hosted in a particular area, for example Olympic games, or political elections, which can call for a temporary needed increase in network capacity, the ability of UAVs network to be quickly deployed can supply this much needed capacity increase. The UAV network can in fact act as a back-haul for whatever type of connection is needed, both cellular data or WLAN, guaranteeing a flexible way to provide momentary and on demand relief for terrestrial networks.

\subsection{Target-specific CDN and Caching}
Another pretty useful aspect of having a mobile UAV network is that of using UAVs and Content Delivery Network nodes. Wireless data traffic has been increasing in recent years due to the greater demand of content-centric communications (such as video and music streaming). This determines a heavy burden in current back-haul links for this type of content, plus, since mobile users are constantly moving, a more flexible caching strategy is preferred. Thus, using UAVs BTS can be used to dynamically cache the popular contents, track the mobility patterns of wireless devices and then effectively serve them, reducing transmission latency and alleviating the traffic load, especially on back-haul links. In UAV-assisted caching, the more interesting contents can be directly cached in the UAV-BSs and then distributed to users.

\subsection{Target-specific Edge Computing}
Similar to this application is that of using UAVs BTS as edge computing devices to offload tasks and relieve computational burdens from the computing server. By adopting a distributed computing architecture with drones as its node, it is in fact possible to move the computation closer to the clients and so reduce network resource demands needed to communicate back and forth between the clients and the required computation node. By adopting a strategy similar to that of the aforementioned CDN/caching purpose, the drone with the required computation node can be kept mobile, instead of fixed as it is now, and shift its position according to which geographical area needs that specific node the most, hence providing lower response time to the final client.
Both of these two techniques will be definitely useful in dramatically increasing the throughput of a network, providing not only relief to the terrestrial network, but also better overall quality of service and quality of experience for the final client.

\section{Conclusion}
The application of UAVs networks are not limited only to the aforementioned scenarios, their actual application possibilities are endless and will evolve even more as newer technologies will push the current boundaries even further. Consider all the possible applications regarding UAV-to-UAV communication for the creation of ad-hoc networks, or even the possibility of directly connecting UAVs to satellites, enabling Satellite-to-UAV communications. With this research we have only scratched the surface, as this is only the beginning of a new generation of networks.

\ifCLASSOPTIONcaptionsoff
  \newpage
\fi

\begin{IEEEbiography}[{\includegraphics[width=1in,height=2in,keepaspectratio]{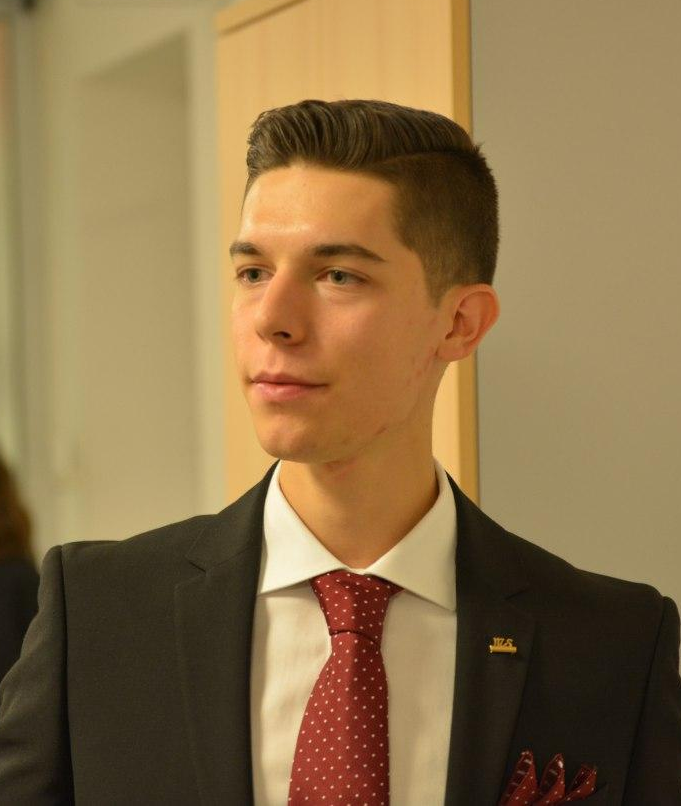}}]{Matteo Strada}
 received his BSc in Computer Science from University of Trento in 2019. He is currently enrolled as MSc student in the EIT Digital Academy, pursuing a double degree program in Cloud and Network Infrastructures studying at University of Trento and TUBerlin.
In 2015 he worked as an intern in Athonet's Srl R\&D department, working on 4G related technologies. From 2016 to 2020 he served as Team Manager and Tech and Sales Consultant for FpvGame 4Front Srl, while at the same time actively competing in FPV and Multirotor drone racing hence acquiring an extensive background in drones and their related technologies.
\end{IEEEbiography}


\begin{thebibliography}{1}
\bibitem{Leonardo}
Leonardo Company, \emph{https://www.leonardocompany.com/it/press-release-detail/-/detail/17-01-2020-leonardo-s-falco-xplorer-drone-completes-first-flight?f=\%2Fpress-release-detail}, 17th January 2020.

\bibitem{Athonet}
Athonet Company, \emph{https://www.athonet.com/missioncritical/}, 17th January 2020.

\bibitem{RCprotocols} 
O.~Liang, \emph{https://oscarliang.com/pwm-ppm-sbus-dsm2-dsmx-sumd-difference/}, February 2018.

\bibitem{UAVComms} 
B.~Li, Z.~Fei and Y.~Zhang, \emph{UAV Communications for 5G and Beyond: Recent Advances and Future Trends}, 20th January 2019.

\bibitem{UAVNetworks}
M.~Mozaffari, W.~Saad, M.~Bennis, Y.~Nam, and M.~Debbah, \emph{A Tutorial on UAVs for Wireless Networks: Applications, Challenges, and Open Problems}, 17th March 2019.

\bibitem{Songbird}
H.~Thamm, N.~Brieger, K.~Neitzke, M.~Meyer, R.~Jansen and M.~Mönninghof, \emph{SONGBIRD - an innovative UAS combining the advantages of fixed wing and multi rotor UAS}, 30th August 2015.

\bibitem{GasMultirotor}
T.~Pang, K.~Peng, F.~Lin and B.~Chen, \emph{Towards Long-endurance Flight: Design and Implementation of a Variable-pitch Gasoline-engine Quadrotor}, 1st June 2016.

\bibitem{SystemDowntime}
B.~Felter, \emph{https://www.vxchnge.com/blog/ways-system-downtime-could-impact-company}, 10th May 2019.

\end{thebibliography}
\end{document}